\documentclass[a4paper,11pt]{article}
\usepackage{latexsym}	
\usepackage{amsmath}
\usepackage{amsthm}

\usepackage{amsmath}
\usepackage{amsfonts}
\usepackage{amssymb}
\usepackage{amsthm}

\usepackage{graphicx}
\usepackage{xcolor}
\usepackage{url}

\usepackage[english]{babel}
\usepackage[utf8]{inputenc}

\newcommand{\R}{{\mathbb R}}
\newcommand{\C}{{\mathbb C}}
\newcommand{\N}{{\mathbb N}}
\newcommand{\Z}{{\mathbb Z}}
\newcommand{\Q}{{\mathbb Q}}

\theoremstyle{plain}

\numberwithin{equation}{section}

\theoremstyle{plain}

\begin{document}
\title{Exact quantum revivals for the Dirac equation}
\author{Fernando Chamizo
\and
Osvaldo P. Santill\'an}
\date{December 20, 2023}
\pagestyle{plain}

\maketitle

\begin{abstract}
In the present work, the results obtained in \cite{strange} about the revivals of a relativistic fermion wave function on a torus are considerably enlarged. In fact, all the possible quantum states exhibiting revivals
are fully characterized. The revivals are exact, that is, are true revivals without taking any particular limit such as the non relativistic one.  The present results are of interest since they generalize the Talbot effect and the revivals of the Schr\"odinger equation to a relativistic situation
with non zero mass. This makes the problem nontrivial, as the dispersion relation is modified and is not linear. The present results are obtained by the use of arithmetic tools which are
described in certain detail. In addition, several plots of the revivals are presented, which are useful for exemplifying the procedure proposed along the text.
 \end{abstract}

\section{Introduction}

One of the first historical situations when the revivals of an initial state was of particular interest is the Talbot effect.  This effect appears when a plane electromagentic wave is incident upon a periodic diffraction grating, the image of the grating is repeated at regular distances away from the grating plane \cite{oskolkov}-\cite{ToRo}. This effect of course, is related to wave equations, which are non dispersive.

The existence of revivals in the context of Quantum Mechanics is instead more surprising.
The  Schr\"odinger equation is an example of a dispersive equation, and it implies that when time evolves  the solution is expected to randomize or spread out. This can be considered as a manifestation of the uncertainty principle. A basic example is the evolution of an initial Gaussian packet, which at a  time $t$ is given by
\[
 \Psi(\mathbf{r},t)=
 \frac{\exp\big(-\frac 12(a+i\hbar t/m)^{-1}\mathbf{r}\cdot\mathbf{r}\big)}{(\pi a)^{3/4}(1+i\hbar t/ma)^{3/2}}.
\]
Despite this dispersive behavior, in some situations there is a coherence between the phase velocities causing \emph{quantum revivals}. This is an exact or approximate periodicity in time. For instance, if the energy spectrum is discrete, say $\{E_n\}_{n=1}^\infty$, and there exists $s\in\R^+$ such that $\{sE_n\}_{n=1}^\infty\subset\Z$ then the solutions of the Schr\"odinger equation are of the form 
\[
 \Psi(\mathbf{r},t)=
 \sum_{n=1}^\infty
 a_n \psi_n(\mathbf{r})\,e^{-\frac{iE_n t}{\hbar}},
\]
and an exact revival $\Psi(\mathbf{r},t)=\Psi(\mathbf{r},t+T_{\text{\rm rev}})$ appears, 
with $T_{\text{\rm rev}}=\frac{2\pi s}{\hbar}$ being the \emph{revival time}. 
A simple example  of this situation is the infinite quantum well. The changes suffered by the initial state at a fractional multiple of the revival time and its fractalization for irrational multiples have been studied for several quantum systems in many works 
\cite{berry}, \cite{olver}, \cite{oskolkov}, \cite{eceizabarrena}, \cite{HaLo}, \cite{ChSa}, 
establishing some interesting connections with the classic Talbot effect in optics and certain topics in number theory \cite{BeKl}. 
\medskip 

There exists simulations related to quantum particles  in boxes \cite{marzoli}, \cite{marzoli2} which may might suggest that, by taking into account relativistic effects, the coherence of the phase velocities is lost and one can only expect approximate quantum revivals (as it occurs in many models cf. \cite{ToRo}). However,  in \cite{strange} it is shown a different truth. In this reference, a fermionic Dirac particle confined to a circle of radius $R$ is studied,  and it is shown that if the dimensionless quantity $q=\frac{McR}{\hbar}$, where $M$ is the mass, is a positive even integer, then one can construct a finite number of plane wave spinor solutions presenting exact quantum revivals. An elegant part of the argument is that these plane waves are associated to Pythagorean triples. These are integral solutions of the algebraic equation  $x^2+y^2=z^2$ \cite[\S13.2]{HaWr}) which constitute a very old topic in pure mathematics. This observation introduces a new connection between quantum revivals and number theory, to be explored further in the present paper.
\medskip

With the methods developed in  \cite{strange}, it follows that for  $q\in 2\Z^+$ there are only few possible plane wave solutions which exhibit quantum revivals. These solutions are related to the divisors of $q$ (on average this number goes like $\log q$ \cite[\S18.2]{HaWr}). This fact suggests that there is little freedom to construct examples following those lines.  In this paper this model is analyzed  with more advanced arithmetic tools. In these terms, it turns out that for each~$q$ satisfying a less restrictive condition, apart from some trivial cases, there are infinitely many plane wave solutions such that any linear combination of them presents exact revivals. 
In fact, for each valid $q$ all the states showing these revivals are characterized in an algorithmic way. 
The exponential growth of the energies explains some fractal like  \emph{quantum carpets}\footnote{Which are density plots of the state along a period \cite{BeMa}.} and connects with some topics in fractal geometry.
The case of particles confined in a square flat torus is also studied. The number of plane waves serving as building  blocks  to obtain the states possessing exact quantum revivals is still infinite for suitable fixed values of $q$ in a dense set, but in this 2-dimensional case the possible energies present a mild growth allowing high degeneracies and more freedom to choose the initial conditions.  This will be described in detail along the text, and involves an interesting connection with Fuchsian groups.

The present work is organized as follows. In section 2 the defining equation and the representation employed along the text are clarified.
Also, the main problem of characterization of revivals is stated. In section 3 the possible revivals in a one dimensional torus are characterized in terms
of solutions of Pell's equations, and the presence of an infinite number of states exhibiting quantum revivals is pointed out.
In section 4 the two dimensional case is analyzed. This situation is more complex and its analysis relies on arithmetic topics related to Fuchsian groups.  Section 5 contains several examples with numerical simulation, which are aimed
for clarifying the main procedure described along the text.
Section 6 contains the discussion of the presented results.

\section{The basic model}

In the following, the plane wave solutions of the Dirac equation in three dimensions
$i \gamma^\mu \partial_\mu \Psi-m \Psi=0,$
under the representation $\gamma^0=\sigma^0$, $\gamma^1=i\sigma^1$ and $\gamma^2=i\sigma^2$ with $\sigma^i$ the standard Pauli matrices, will be considered.
The particles will be assumed to be confined in a flat torus obtained identifying the opposite sides of 
$ [2\pi R_1, 0]\times [0, 2\pi R_2]$. In this situation the cartesian coordinates may be parameterized as $x=R_1\phi_1$, $y=R_2\phi_2$, with $\phi_1$ and $\phi_2$ two $2\pi$-periodic angles. This corresponds
to periodic boundary conditions on the wave function,  which correspond to relativistic Bloch states $u \exp({-\frac{ik\cdot x}{\hbar}})$. It imposes the following  quantization of the momentum 
$$
 \vec{k}=
 \Big(
 \frac{\hbar n_1}{R_1},
 \frac{\hbar n_2}{R_2}
 \Big)
 \qquad\text{with}\quad
 \vec{n}=(n_1,n_2)\in\Z^2. 
$$
Let $\omega_{\vec{n}}=\hbar^{-1}E_{\vec{n}}>0$ be the  angular frequency for each $\vec{n}$ and consider the complex quantity
$$z_{\vec{n}} = 
\frac {\hbar n_1}{R_1}
 +
 \frac{i \hbar n_2}{R_2},$$
associated to the momentum.  The generic energy equation $E^2 = M^2c^4+c^2\vec{k}^2$ implies 
$$
\omega_{\vec{n}}
 =
 \frac{c}{\hbar}
 \sqrt{M^2c^2+|z_{\vec{n}}|^2},
$$
which in particular gives the following useful relation 
$$
 \Big(
 \frac{\hbar \omega_{\vec{n}}}{c}+Mc
 \Big)^2
 +|z_{\vec{n}}|^2
 =
 \frac{2\hbar\omega_{\vec{n}}}{c}
 \Big(
 \frac{\hbar\omega_{\vec{n}}}{c}+Mc 
 \Big).
$$
The normalized solution is in these terms
\[
 \Psi_{\vec{n}}(\phi_1,\phi_2,t)
 =
 \sqrt{\frac{E_{\vec{n}}+Mc^2}{2E_{\vec{n}}}}
 \begin{pmatrix}
  \frac{z^\ast_{\vec{n}}}{E_{\vec{n}}+Mc^2}
  \\[5pt]
  1
 \end{pmatrix}
 e^{i(n_1\phi_1+n_2\phi_2-\omega_{\vec{n}}  t)}.
\]
We restrict ourselves to the fermionic case since similar considerations holds for antifermions. 

Some particular limits are in order. The 1-dimensional case corresponds to let $R_1\to\infty$, so the periodicity in the $x$ coordinate is killed, $k^1=0$ and the particle is confined to move in a (flat) circle of radius $R=R_2$ given by  a $2\pi R$-periodic $y$ coordinate. 
The 2-dimensional case corresponds to the model above choosing $R=R_1=R_2$ (a square flat  torus). 
Therefore, $z_{\vec{n}} = \frac{i\hbar n_2}{R}$ in the first case and $z_{\vec{n}} = \frac{\hbar( n_1+in_2)}{R}$  in the second one. 

Both cases above corresponding to $R_1=R_2=R$ can be written in a unified way by introducing the dimensionless positive quantity
$ q=\frac{McR}{\hbar}$, from where
 $ \omega_{\vec{n}}=\frac{c}{R}\sqrt{\vec{n}^2+q^2}$, and the formula
\begin{equation}\label{estate}
 \Psi_{\vec{n}}(\phi_1,\phi_2,t)
 =
 \frac{1}{\sqrt{2}} \bigg(
1+\frac{q}{\sqrt{\vec{n}^2+q^2}}
\bigg)^{\frac{1}{2}}
 \begin{pmatrix}
  \frac{n_1-in_2}{q+\sqrt{\vec{n}^2+q^2}}
  \\[5pt]
  1
 \end{pmatrix}
 e^{i(n_1\phi_1+n_2\phi_2-\omega_{\vec{n}}  t)}.
\end{equation}
follows. The 1-dimensional case corresponds to set $n_1=0$ (so, the state does not depend on  $\phi_1$) and the $2$-dimensional case to leave it free. 
The  $\frac{1}{2\pi}$ factor was taken apart from the normalization in order to avoid conflicts with the choice of the dimension. 

In order to make a comparison with \cite{strange} it may be mentioned that this author considers a wave function with four components, and the corresponding $\gamma^\mu$
matrices are of order $4\times 4$.  This is different from the present case. However the revival issues, which are of central importance in the present paper, is only sensitive
to the phase and not to the normalization factors or the number of components of the spinor.

The spinor given above is an energy eigenfunction. A generic state  having a discrete, finite and nonnegative energy spectrum is  a superposition of the form
\begin{equation}\label{state}
 \Phi = 
 \sum_{\vec{n}\in\mathcal{N}} c_{\vec{n}} \Psi_{\vec{n}}
 \qquad\text{with}\quad c_{\vec{n}}\in\C-\{0\}
\end{equation}
where $\mathcal{N}=\big\{\vec{n}_0,\vec{n}_1,\dots, \vec{n}_N\big\}$ contains $N$ vectors, 
$\vec{n}_j=(k_j,l_j)\in\Z^2$ and $k_j=0$ in the one dimensional model. The energy spectrum is related to  the squared norm of the $\vec{n}_j$. It can be described as the set
$$
 \mathcal{E}
 =
 \Big\{
 \frac{c\hbar}{R}
 \sqrt{k_j^2+\ell_j^2+q^2}\,:\, (k_j,\ell_j)\in\mathcal{N}
 \Big\}.
$$
The cardinality $\#\mathcal{E}$ of this set may be different from $N$, as there is a possible degeneracy of states with the same energy value.
At this point, it is important to state the following affirmation, which is a direct and simple consequence of the spinor formulas given above.

\begin{quote}
 \emph{Proposition 1:} The state described by the formulas \eqref{estate}--\eqref{state} is periodic in time, and thus exhibits quantum revival, if and only if for each $0\le j\le N$ there exists an irreducible fraction $\frac{a_j}{b_j}>0$ such that 
 \begin{equation}\label{cond}
  k_j^2+\ell_j^2+q^2=\frac{a_j^2}{b_j^2}\big(k_0^2+\ell_0^2+q^2\big).
 \end{equation}
 Here $N$ is the number of elements of the state \eqref{state}. The revival time is given by the formula
 \begin{equation}\label{lcm}
  T_{\rm rev}
  =\frac{2\pi L}{\omega_{\vec{n}_0}}
  \qquad\text{with}\quad
  L=\text{\rm lcm}\, \big(\{b_j \}_{j=0}^N\big),
 \end{equation}
 where, as usual, $\text{\rm lcm}$ denotes the least common multiple.
\end{quote}

\emph{Proof:} This proposition follows directly from the formulas just stated.  In fact, consider a generic wave function \eqref{state} expanded in terms of the eigenfunctions \eqref{estate}.
For fixed values of $\phi_1$ and $\phi_2$, recalling the formula for $\omega_{\vec{n}}$,  there will be a revival if for all the $0\le j\le N$ there is a time $T_{\rm rev}$, the revival time, such that
$$
\sqrt{\vec{n}_j^2+q^2}\,c T_{\rm rev}=2\pi R m_j, \qquad m_j\in \Z
$$
The time like phase of the state \eqref{state} in this case will be a multiple of $2\pi$ and will reproduce the initial wave function at $t=0$.
Now, as the last formula follows from every $j$, it follows that 
$$
(\vec{n}_j^2+q^2)=\frac{m_j^2}{m_0^2}(\vec{n}_0^2+q^2).
$$
This is of the form \eqref{cond} after simplifying the common factors between $m_0$ and $m_j$. 
We have 
$\frac{\omega_{\vec{n}_0}}{2\pi} T_{\rm rev}= m_j \frac{\omega_{\vec{n}_0}}{\omega_{\vec{n}_j}}=m_j\frac{b_j}{a_j}$.
As $m_j$ is a multiple of $a_j$ and $T_{\rm rev}$ is defined as the minimal period, then $\frac{\omega_{\vec{n}_0}}{2\pi} T_{\rm rev}$ equals $L$, as stated in \eqref{lcm}. (Q. E. D)

\
 
 The revival problem has been reduced to the problem of solving equation \eqref{cond}. The corresponding solutions are not necessarily trivial from the arithmetic point of view.
In fact, as it will be shown below, they are described in terms of generalized Pell's equations in the one dimensional case and they relate to some Fuchsian groups in the 2D case.

The philosophy that will guide the present paper is perhaps different than the standard one. Instead of looking for particular states which exhibit quantum revivals,
the intention is to characterize all of them and it is achieved fixing a specific vector state $\vec{n}_0$ and considering all possible $\Phi$ in \eqref{state} containing $\Psi_{\vec{n}_0}$.

\begin{quote}
\emph{Philosophy:} Consider a pivot eigenstate $\Psi_{\vec{n}_0}$.  The idea is to
characterize the possible sets $\mathcal{N}$ containing $\vec{n}_0$ such that  quantum revivals appear. More concretely, given $\vec{n}_0$ we want to determine the maximal set of quantum numbers $\mathcal{N}_0$ including $\vec{n}_0$ such that 
the wave function
\begin{equation}
\Phi =  \sum_{\vec{n}\in\mathcal{N}}  c_{\vec{n}} \Psi_{\vec{n}}
 \end{equation}
verifies
\begin{equation}\label{maxi}
 \boxed{
 \Phi\text{ is periodic in }t
 \quad\Leftrightarrow\quad 
 \mathcal{N}\text{ is a finite subset of }\mathcal{N}_0.
 }
\end{equation}
\end{quote}

 Note that, once the revival set $\mathcal{N}_i$ for an arbitrary pivot eigenstate $\Psi_{\vec{n}_i}$ is determined, given a composite wave function 
$$
\Phi_c= 
 \sum_{i} c_{i} \Psi_{\vec{n}_i},
 $$
 its most general revival wave function is 
$$
\Phi_r= 
 \sum_{\vec{n}\in\mathcal{N}_I} c_{\vec{n}} \Psi_{\vec{n}},
 $$
 where $\mathcal{N}_I$ is the intersection of all the sets $\mathcal{N}_i$.
 
 In addition, it is not difficult to see that $\mathcal{N}_0$ is not uniformly bounded. This can be exemplified even in the simplest case, in which one forces all the states to have the same energy, that is, 
  $\#\mathcal{E}=1$. In the one dimensional situation $\Psi_{(0,\ell_0)}$ can be only accompanied with $\Psi_{(0,-\ell_0)}$ giving obvious revivals in \eqref{state} and $\mathcal{N}_0=2$. This situation is uninteresting. However, in the 2D case the situation is less trivial because $\#\mathcal{N}$ can be arbitrarily large keeping $\#\mathcal{E}=1$ because the function giving the number of representations as a sum of two squares is unbounded\footnote{Although it is $\pi$ on average, as it is easily checked approximating the number of integer points  in a circle by its area \cite[Th.\,339]{HaWr}.}. So, $\mathcal{N}_0$ is finite but not uniformly bounded in {terms} of $\vec{n}_0$. For instance, $\vec{n}_0=(178,19)$ can be complemented with other $63$ integral pairs to get the same 
$ \frac{R\omega_{\vec{n}_j}}{c}=\sqrt{\vec{n}_0^2+q^2}=\sqrt{32045+q^2}$. Example of these pairs are
\[
        178^2 + 19^2 =
        166^2 + 67^2 =
        157^2 + 86^2 =
        179^2 + 2^2 =
        142^2 + 109^2 =
        163^2 + 74^2 =
        \dots
\]
This phenomenon of unbounded degeneracy of the energy leads typically to intricate density probabilities for a fixed time (see the section of numerical results) and it is linked to some unsolved problems about the structure of the nodal lines \cite{BoRu}, \cite{berry_nodal}. 

The remaining part of the present work is aimed to show the complexity of the set $\mathcal{N}_0$, without restricting the attention to just one energy eingenvalue.

\section{The one dimensional case}

Recall that in the one dimensional case the first coordinate of $\vec{n}$ is set to $0$, and the necessary and sufficient condition \eqref{cond} for having quantum revivals becomes
\begin{equation}\label{1dcond}
 \ell_j^2+q^2
 =
 \frac{a_j^2}{b_j^2}
 \big(\ell_0^2+q^2\big)
 \qquad\text{for}\quad 0\le j\le N.
\end{equation}
Since $l_0$, $l_j$, $a_j$ and $b_j$ are integers, it is clearly seen that $q^2$ should be rational. In fact, if $q^2\not\in\Q^+$ then $\frac{a_j}{b_j}=1$ and $\#\mathcal{E}=1$ leading to a trivial case in which the wave function is composed by two eigenstate
corresponding to $l_0$ and $-l_0$, as mentioned before. Therefore,  it can be assumed safely that $q^2\in\Q^+$. In fact, it is convenient to focus on $q^2\in\Z^+$ because it is somewhat simpler and contains all the ingredients appearing in the general situation.

Some concepts of arithmetic are in order. First, the standard number theory notation $a|b$ will be employed,  to express that the integer $a$ divides
the integer $b$. In addition, recall that a positive integer is said to be \emph{squarefree} if it is not divisible by  a square different from~$1$. Each positive integer $m$ admits a unique factorization as a product of a squarefree and a square, in the form $m=D r^2$
such that $\sqrt{D}\notin \Q$ and $r\in \Z$. Consider such decomposition for $\ell_0^2+q^2$:
\begin{equation}\label{sqf}
 \ell_0^2+q^2=Ds^2,
 \end{equation}
 with $D$ squarefree.
Then \eqref{1dcond} can be arranged as 
\begin{equation}\label{pellj}
 \ell_j-D
 \Big(\frac{a_js}{b_j}\Big)^2=-q^2.
\end{equation}
As $D$ is squarefree and $q^2$ is integer, $\frac{a_js}{b_j}$ is integer too. So, all the possibilities for $\ell_j$ and the corresponding $\frac{a_j}{b_j}$ are in one to one correspondence with the integer solutions of the \emph{generalized Pell's equation}
\begin{equation}\label{gpell}
 x^2-Dy^2=-q^2.
\end{equation}
In these terms, the Proposition 1 of the previous section, adapted to the one dimensional case, is the following.

\begin{quote}
\emph{Proposition 2:} Given $\ell_0\in\Z$ and $q^2\in\Z^+$, defining $D$ and $s$ as in \eqref{sqf}, the set $\mathcal{N}_0$ of quantum numbers to have \eqref{maxi} when $(0,\ell_0)\in\mathcal{N}$ is
\begin{equation}\label{novoa1}
  \mathcal{N}_0
  =
  \big\{
  (0,x)\,:\, x^2-Dy^2=-q^2\text{ with }(x,y)\in\Z^2.
  \big\}.
 \end{equation}
 Moreover the revival time is given by 
 \[
  T_{\rm rev}
  =
  \frac{2\pi R L}{cs \sqrt{D}}
  =
  \frac{2\pi L}{\omega_{\vec{n}_0}},
 \]
 where $L$ is the $\text{\rm lcm}$ of the denominators of $$\frac{a_j}{b_j}=\frac1s {\sqrt{ \frac{\ell_j^2+q^2}D }},$$ for the $(0,\ell_j)\in\mathcal{N}.$
\end{quote}

In view of the last proposition, it is mandatory to review the solutions of the generalized Pell's equation \eqref{gpell}
that allows to describe the structure of $\mathcal{N}_0$ for any choice of $\ell_0\in\Z$ and $q^2\in \Z^+$. 

The case $D=1$ is somewhat trivial because $x^2-Dy^2=(x-y)(x+y)=-q^2$ implies that $x-y$ and $x+y$ are divisors of $q^2$ leaving only a finite number of possibilities. In other words, if $\ell_0^2+q^2$ is a square, $\mathcal{N}_0$ is finite. This includes the case covered in \cite{strange} using Pythagorean triples. 
If $D\ne 1$, $\mathcal{N}_0$ is instead an infinite set, which can be seen as follows. For the standard \emph{Pell's equation}
\begin{equation}\label{hold}
  x^2-Dy^2=1,
\end{equation}
a classic result due to Lagrange is of fundamental importance for finding the solution.

\begin{quote}
\emph{Lagrange algorithm:} If the constant $D\in\Z^+$ is not a square (this is assured in the present case since $D$ is squarefree and $D\ne 1$) then the equation \eqref{hold} always has infinitely many solutions $(x,y)\in\Z^+\times \Z^+$. These can be expressed  in the synthetic way as follows
\begin{equation}\label{recurri}
  x+y\sqrt{D}=\big(x_p+y_p\sqrt{D}\big)^n
  \qquad\text{with}\quad n\in\Z^+
\end{equation}
where $(x_p,y_p)$ is the minimal positive solution. Namely, the one having $x_p$ and hence $y_p$ as small as possible. 
\end{quote}

It can be checked directly that the previous formula gives solutions by simply multiplying by the conjugate because 
$$
 \big(x_p+y_p\sqrt{D}\big)\big(x_p-y_p\sqrt{D}\big)
 =
 x_p^2-Dy_p^2=1.
$$
In the present case, it is deduced from \eqref{sqf} that $(\ell_0,s)$ is an integer solution of \eqref{gpell}. The same argument just employed, using the conjugate, shows that
\begin{equation}\label{gpf}
  x+y\sqrt{D}
  =
  \pm
  \big(\ell_0+s\sqrt{D}\big)
  \big(x_p+y_p\sqrt{D}\big)^n
  \qquad\text{with}\quad n\in\Z
\end{equation}
are infinitely many integer solutions of \eqref{gpell}, proving that $\mathcal{N}_0$ is infinite. The $\pm$ and the extension of the range of $n$ from $\Z^+$ to $\Z$ are only to take into consideration different combinations of signs. Note that $x_p-y_p\sqrt{D}=\big(x_p+y_p\sqrt{D}\big)^{-1}$.
\medskip 

By defining the $n$-th family of solutions as $x_n+y_n\sqrt{D}=\big(x_p+y_p\sqrt{D}\big)^n$, it follows that 
 \eqref{recurri}  can be written as 
 $$
  x_{n+1}+y_{n+1}\sqrt{D}=\big(x_p+y_p\sqrt{D}\big)(x_n+y_n \sqrt{D}).
 $$
 This leads to the first order recurrence
 $$
 x_{n+1}=x_n x_p+D y_n y_p, \qquad y_{n+1}=x_p y_n+x_n y_p.
 $$
 By repeating this step once more leads to an expression for $x_{n+2}$ as a linear combination of $x_n$ and $y_n$. So  is $x_{n+1}$. By eliminating the $y_n$ variable with this two equations, it is obtained a second order linear
 recurrence for  $x_n$ of the generic form
 $$
 x_{n+2}=\alpha x_{n+1}+\beta x_n,
 $$
 with $\alpha$ and $\beta$ constants depending on $D$, $x_p$ and $y_p$.
 The variable $x_n$ is the fundamental quantity describing $\mathcal{N}_0$. A concrete example is given in the following sections, more precisely in the formula \eqref{nina}.

A complication of the theory (related to deep topics in number theory as unique factorization and the class number \cite{cohn})
is that a finite number of similar families of solutions may exists. Namely, in our case they may appear families of integer solutions replacing $\ell_0$ and $s$ by $a$ and $b$ with $(x,y)=(a,b)$ satisfying \eqref{gpell} but it can be proved that $0<a<q(1+\sqrt{2x_p})/2$
\cite{conrad}
then there is only a finite number of possibilities to explore.
All the possible families obey to the same recurrence law because they always come from the multiplication by $x_p+y_p\sqrt{D}$.

Note that always \eqref{gpf} can be translated in one sided recurrence formulas dividing into $n\ge n_0$ and $n<n_0$. With this remark and the previous considerations, it is deduced the following.

 \begin{quote}
 \emph{Proposition 3:} There exists a finite number of sequences $c_n^{(1)}$,\dots, $c_n^{(J)}$ satisfying a second order linear recurrence such that 
 \[
  \mathcal{N}_0=
  \big\{ (0,\pm c_n^{(j)}) \text{ for {$1\le j\le J$} and $n\ge 0$}\big\}.
 \]
\end{quote}

Let us illustrate the situation with an example. If $\ell_0=5$ and $q=\sqrt{2}$ then \eqref{sqf} produces $D=3$ and $s=3$. A direct search shows that $(x_p,y_p)=(2,1)$ is the minimal positive solution of $x^2-3y^2=1$. The values $n=-1,0,1,2$ in \eqref{gpf} give 
\[
 x+y\sqrt{3}
 =
 \pm(1+\sqrt{3}),\, 
 \pm(5+3\sqrt{3}),\, 
 \pm(19+11\sqrt{3}),\, 
 \pm(71+41\sqrt{3}). 
\]
Hence $(0,\pm 1)$, $(0,-5)$, $(0,\pm 19)$ and $(0,\pm 71)$ are in the $\mathcal{N}_0$ corresponding to $(0,\ell_0)=(0,5)$.
For this example, it can be proved (see the section of numerical results) 
\begin{equation}\label{example1d}
  \mathcal{N}_0
 =
 \big\{
 (0,\pm c_n)\text{ where }
 c_{n+2}=4c_{n+1}-c_n
 \text{ with }
 c_0=1,\ c_1=5
 \big\}.
\end{equation}
There are not more families because the bound $0<a<q(1+\sqrt{2x_p})/2$ only allows $(a,b)=(1,\pm 1)$ which belong to the same family as $(\ell_0,s)=(5,3)$ because
$1+\sqrt{3}=(5+3\sqrt{3})(2+\sqrt{3})^{-1}$ and
$1-\sqrt{3}=-(5+3\sqrt{3})(2+\sqrt{3})^{-2}$.

The sequence, due to \eqref{gpf},  has an exponential growth
\[
 \{c_n\} = 
 \{1,5,19,71,265,989,3691,13775, 51409 , 191861 , 716035 , 2672279 , 9973081 , \dots\}.
\]
Depending on the elements chosen to compose $\mathcal{N}_0$, the revival time can be at most $\frac{6\pi}{\omega_{\vec{n}_0}}$, because $\frac{a_js}{b_j}=\frac{3a_j}{b_j}$ must be an integer and hence $L$ divides $3$.
\medskip

For an exhaustive description of the generalized Pell's equation, some interesting references are \cite{shockley}, \cite{conrad}, \cite{cohn} and \cite{JaWi}. 
\medskip

The next task is to mention how the previous ideas can be modified to cover the remaining cases $q^2\in\Q^+-\Z$ preserving the same result. This will be done briefly.
In this situation, \eqref{sqf} is substituted by the squarefree-square decomposition of the numerator and the denominator of $\ell_0^2+q^2$ to get
\[
  \ell_0^2+q^2=\frac{Ds^2}{D_*s_*^2}
 \qquad\text{with $D$ and $D_*$ squarefree}. 
\]
Clearing denominators in \eqref{pellj}, it is obtained that $(\ell_j, \frac{a_j s}{b_j})$ is a solution of 
\[
 D_* s_*^2x^2-Dy^2=-q^2D_*s_*^2.
\]
In fact it is an integral solution because $-q^2D_*s_*^2\in\Z$ and the tentative denominator of $(\frac{a_js}{b_j})^2$ could not be canceled with $D$ because the latter is squarefree. So, to cover the full rational case in the affirmation above, $\mathcal{N}_0$ must be generalized to 
\begin{equation}\label{n01dr}
  \mathcal{N}_0
  =
  \big\{
  (0,x)\,:\, D_*s_*^2x^2-Dy^2=-q^2D_*s_*^2\text{ with }(x,y)\in\Z^2
  \big\}.
\end{equation}
Note that this is actually a generalization because $D_*s_*^2=1$ if $q^2\in\Z$.
\medskip 

The case $D=D_*=1$ leads, as before, to a finite number of possibilities because $s_*x+y$ and $s_*x-y$ are divisors of $-q^2s_*^2$.
The explicit description of $\mathcal{N}_0$ is
\[
 \mathcal{N}_0
 =
 \Big\{
 \Big( 0, \frac{q^2s_*^2-d^2}{2s_*d}\Big)\,:\, d\mid q^2s_*^2\text{ and }\frac{q^2s_*^2}{d}-d\text{ multiple of }2s_*
 \Big\}.
\]
When $s_*$ grows, being a multiple of $2s_*$ imposes a strong condition and $\mathcal{N}_0$ might reduce to the trivial set $\{(0,\pm \ell_0)\}$ even if $q^2s_*^2$ has nontrivial divisors. An example of this situation is $\ell_0=2$, $q=\frac 25\sqrt{299}$. 
\medskip 

In the rest of the cases ($D$ and $D_*$ not simultaneously~$1$), a modification of the arguments for $q^2\in\Z$ applies. Recall that $(\ell_0,s)$ is a solution of the equation in \eqref{n01dr}. Reasoning as in \eqref{gpf} multiplying by the conjugate, it is deduced that
\[
 x s_*\sqrt{D_*}+y s\sqrt{D}
 =
 \pm
 \big(\ell_0s_*\sqrt{D_*}+s\sqrt{D}\big)
 \big(x_p+y_pss_*\sqrt{DD_*}\big)^n
 \qquad\text{with}\quad n\in\Z
\]
gives infinitely many integer solutions  where $(x_p,y_p)$ is the minimal positive solution of the Pell equation 
\[
 x^2-s^2s_*^2DD_*y^2=1.
\]
 Again, a finite number of other families may appear.

Summing up, for $q^2\in\Q$ the set $\mathcal{N}_0$ is infinite except in the case in which $\ell_0^2+q^2$ is the square of an irreducible fraction.

\section{The two dimensional case}

The two dimensional  case is much more involved because, as it will be shown below, the families of eigenstates presenting exact quantum revivals are parametrized Fuchsian groups i.e., discrete subgroups of $\text{\rm PSL}_2(\R)$.
Recall that \eqref{cond} implies $q^2\in\Q$ except in the case $\#\mathcal{E}=1$ in which $\frac{a_j}{b_j}=1$. As in the one dimensional case, the focus will be on the case $q^2\in\Z$ and some closing comments about the fractional case will be made later on. Some considerations about $\#\mathcal{E}=1$ were included above and it will be expanded in the next section.
\medskip

Given $q^2\in\Z$, consider the decomposition
\begin{equation}\label{sqf2}
 k_0^2+\ell_0^2+q^2=Ds^2
 \qquad\text{with $D$ squarefree}.
\end{equation}
The condition \eqref{cond} reads $$k_j^2+\ell_j^2+q^2=D\bigg(\frac{a_js}{b_j}\bigg)^2,$$ where $\frac{a_js}{b_j}$ must be integer\footnote{This is because $D$ is squarefree and $k_j^2,\ell_j^2,q^2\in\Z$.} and it follows a natural analogue of the one dimensional result.

\begin{quote}
\emph{Proposition 4:}  Given $(k_0,\ell_0)\in\Z^2$ and $q^2\in\Z^+$, then by defining $D$ and $s$ as in \eqref{sqf2}, the set $\mathcal{N}_0$ of quantum numbers needed for  \eqref{maxi} with $(k_0,\ell_0)\in\mathcal{N}$ to hold is
 \[
  \mathcal{N}_0
  =
  \big\{
  (y,z)\,:\, Dx^2-y^2-z^2=q^2\text{ with }(x,y,z)\in\Z^3
  \big\}.
 \]
 Moreover the revival time is given by
 $$
  T_{\rm rev}
  =
  \frac{2\pi R L}{cs \sqrt{D}}=
  \frac{2\pi L}{\omega_{\vec{n}_0}},
$$
  where $L$ is the $\text{\rm lcm}$ of the denominators of $\frac1s {\sqrt{ \frac{k_j^2+\ell_j^2+q^2}D  }}$ for the $(k_j,\ell_j)\in\mathcal{N}$.
\end{quote}

This statement suggests that the study of the group that leaves invariant the quadratic form $Q_{2D}=Dx^2-y^2-z^2$ and which applies integer solutions into integer solutions is of  importance, as it leads to an algorithm generating new solutions.
The resulting group is related to the Lorentz group, but it is rather clear that it does not contain all the elements. The task is to determine which elements are to be included.
This group is not present in the one dimensional case. 

It may be convenient to explain the difference between the one and two dimensional case from an abstract point of view.  As reviewed above, the structure of the solutions in the one dimensional case can be summarized saying that there were finitely many families of solutions and each family was given by the action of the discrete group $\big\{\big(x_p+y_p\sqrt{D}\big)^n\big\}_{n\in\Z}$, which is isomorphic to $\Z$, producing sparse solutions because of the exponential growth. In the 
two dimensional case instead, the group is non-abelian and gives a denser set of solutions. 

In order to study the above mentioned group, let us start with the case $D=1$, which is no longer trivial.
It is known that, see for instance \cite{duke}, that the map
\begin{equation}\label{isom1}
 \begin{pmatrix}
  a&b \\ c&d
 \end{pmatrix}
 \longmapsto
 \begin{pmatrix}
  \frac{a^2+b^2+c^2+d^2}{2} & \frac{-a^2+b^2-c^2+d^2}{2} & -ab-cd
  \\
  \frac{-a^2-b^2+c^2+d^2}{2} & \frac{a^2-b^2-c^2+d^2}{2} & ab-cd
  \\
  -ac-bd & ac-bd & ad+bc
 \end{pmatrix},
\end{equation}
 establishes an isomorphism  between
$\text{\rm PSL}_2(\R)=\text{\rm SL}_2(\R)/\{\pm I\}$ and the proper Lorentz group $\text{\rm SO}^+(1,2)$ of linear transformations leaving invariant the quadratic form $$Q_2(x,y)=x^2-y^2-z^2,$$ up to a global change of sign.
In general, only the matrices in
$\text{\rm SO}^+(1,2)\cap\mathcal{M}_3(\Z)$ with
$\mathcal{M}_3(\Z)$ the set of integral $3\times  3$ matrices, apply
integral solutions into integral solutions.
In the jargon of the quadratic form theory, these matrices are called \emph{integral automorphs}.

Clearly, the fundamental problem is to characterize the preimage of  $\text{\rm SO}^+(1,2)\cap\mathcal{M}_3(\Z)$ by the isomorphism \eqref{isom1} in a simple way. 
The fact that the $3\times 3$ in the right hand of the isomorphism \eqref{isom1} is an integral automorph does not, at first sight, imply that the matrix in the left side has integer entries $a$, $b$, $c$ and $d$. 
The nature of those entries has to be clarified further. First, let $A$ be the matrix in the image of \eqref{isom1}. If $A\in \mathcal{M}_3(\Z)$, as its entries are integers, it is seen that
\[
 2a^2= a_{11}+a_{22}-a_{12}-a_{21}\in\Z.
\]
Other choices of the signs prove that in general $$2a^2, 2b^2,2c^2,2d^2\in\Z.$$ 
On the other hand, using the rest of the entries and the determinant equation $ad-bc=1$,  it is seen that
$$
 2ab=a_{23}-a_{13}\in\Z
 $$
and, similarly,
$$
 2ac,\,2ad,\,2bc,\,2bd,\,2cd\in\Z.
$$
Considering $a_{ij}\pm a_{i'j'}$ with $i,j,i',j'\in\{1,2\}$, it is deduced that $2a^2,\,2b^2,\,2c^2,\,2d^2$ are all even or all odd. For instance $a_{11}+a_{22}$ leads to
$$
\frac{2a^2+2d^2}{2}\in\N,
$$ 
thus $2 a^2$ and $2 d^2$ are simultaneously odd or even. The same line of reasoning applies for the remaining pairs.

The entries $a$, $b$, $c$ and $d$ can be characterized further. Consider first the even case. If one selects a nonzero variable, say $a$, then $2a^2$  is an even integer, thus $a^2$ is an integer. As reviewed in the previous section, 
every positive integer admits a decomposition  as a square free and a square, therefore $a=s_a\sqrt{R_a}$ with $R_a$ an squarefree integer. This last condition means in particular that if $R_a\neq 1$ the number $a$ is not an integer.
On the other hand, as $2ab,2ac,2ad\in\Z$ it follows for instance that
$$
2 s_a s_b \sqrt{R_a R_b} \in \Z.
$$
Here $b=s_b\sqrt{R_b}$ and so on. As $2 s_a s_b$ is an integer, so is $\sqrt{R_a R_b}$. This last requirement implies that $R_aR_b$ is the square of an integer. On the other hand the product of two square free integers $R_a R_b$ gives a square integer only if $R_a=R_b$. After some small reasoning, it follows that $R_a=R_b=R_c=R_d=R$. In other words, $b$, $c$ and $d$ are integers when divided by $\sqrt{R}$. Finally, the determinant equation $ad-bc=1$ implies $R=1$ and it is concluded that $a,b,c,d\in\Z$.

The conditions given above still are not enough. The point is that the quantities in the right hand of \eqref{isom1} should be such that $a_{11},a_{12},a_{21},a_{22}\in\Z$, and this is not necessarily true even when all the conditions above take place. The task is to derive
this further requirement, the result is  that $2\mid a+b+c+d$. This can be seen as follows. 
Note that $\pm n^2-n$ is even for any $n\in\Z$ and any choice of the sign, then $a_{ij}-\frac 12 (a+b+c+d)\in\Z$  for $i,j\in\{1,2\}$, hence $a_{ij}\in\Z$ requires $a+b+c+d$ to be even. 

In these terms,  the searched discrete group is composed by the elements of the sometimes called $\theta$-group
$$
 \Gamma_\theta=
 \left\{
 \begin{pmatrix}
  a&b\\ c&d
 \end{pmatrix}
 \in \text{\rm PSL}_2(\Z)\,:\, 2\mid a+b+c+d
 \right\}.
$$
This covers the case in which $2 a^2$, $2b^2$, $2c^2$ and $2d^2$ are even integers.

 For the odd case instead, one may consider the quantities
$2(a\sqrt{2})^2$, $2(b\sqrt{2})^2$, $2(c\sqrt{2})^2$, $2(d\sqrt{2})^2$ which are clearly even. By repeating the argument it is deduced that
$a\sqrt{2}$, $b\sqrt{2}$, $c\sqrt{2}$, $d\sqrt{2}$ are odd integers. In this case, the elements of the group
\[
C_\theta= \frac{1}{\sqrt{2}}
 \begin{pmatrix}
  1& -1 \\ 1& 1
 \end{pmatrix}
 \Gamma_\theta
\]
are of the searched type since the elements of this coset are of the form
\[
 \begin{pmatrix}
  \frac{(a-c)\sqrt{2}}{2}& \frac{(b-d)\sqrt{2}}{2} \\ \frac{(a+c)\sqrt{2}}{2}& \frac{(b+d)\sqrt{2}}{2}
 \end{pmatrix}
\]
If $a\sqrt{2}$, $b\sqrt{2}$, $c\sqrt{2}$, $d\sqrt{2}$ are odd integers, it is clear that the entries $A_{ij}$ of these matrices are even integers which satisfy $2\mid A_{11}+A_{12}+A_{21}+A_{22}$.
The isomorphism \eqref{isom1} then maps from integer solutions in $\mathcal{N}_0$ one into other.  Conversely, a bit of reasoning shows that any matrix with this property has the form given above. Therefore the isomorphism 
\[
 F:\,
 \Gamma\longrightarrow
 \text{\rm SO}^+(1,2)\cap\mathcal{M}_3(\Z)
\]
holds, where
\begin{equation}\label{grupo}
 \Gamma=
 \Gamma_\theta
 \cup
 C_\theta
 \qquad\text{with}\quad
 C_\theta=
 \frac{1}{\sqrt{2}}
 \begin{pmatrix}
  1& -1 \\ 1& 1
 \end{pmatrix}
 \Gamma_\theta.
\end{equation}
In these terms, the following revival generating algorithm takes place.

\begin{quote}
\emph{Revival algorithm for $D=1$:} Given an element of
$$
\mathcal{N}_0
  =
  \big\{
  (y,z)\,:\, Dx^2-y^2-z^2=q^2\text{ with }(x,y,z)\in\Z^3
  \big\},
$$
with $D=1$,  then \eqref{sqf2} implies that  $x=s$, $y=k_0$, $z=\ell_0$ is a valid solution of $Dx^2-y^2-z^2=q^2$ and it follows that
\begin{equation}\label{sol2d}
 F(\gamma)
 \begin{pmatrix}
  s\\ k_0\\ \ell_0
 \end{pmatrix}
 \qquad\text{with}\quad \gamma\in\Gamma
\end{equation}
gives a family of infinitely many of solutions in $\mathcal{N}_0$.  The group $\Gamma$ is described in \eqref{grupo}.
Generating elements of $\Gamma$ is a simple task because the extended Euclidean algorithm produces easily integer solutions of $aX-bY=1$  (even under a parity condition) to obtain matrices in $\Gamma_\theta$ composed by $a$, $b$, $c=Y$, $d=X$. 
\end{quote}

To give an example of the above procedure, consider the choice $q=\sqrt{6}$, $k_0=1$, $\ell_0=3$ gives $D=1$ and $s=4$ in \eqref{sqf2}. By applying $F(\gamma)$ with $\gamma$ each of the~12 matrices in $\Gamma$ with $|a|,|b|,|c|,|d|\le 2$, there are obtained the  following valid values $(k_j,\ell_j)$ satisfying \eqref{cond}
\[
 \pm (1,3),\quad
 \pm (1,-3),\quad
 \pm (3,-1),\quad
 \pm (3,7),\quad
 \pm (13,9),\quad
 \pm (15,-13).
\]

The case $D>1$ is more difficult although it runs in similar lines. Fortunately,  there exists literature about the subject that dates from more than a century ago \cite{FrKl}.
The initial idea is that the change of variables $x\mapsto \frac{x}{\sqrt{D}}$ passes the quadratic form $$Q_{2D}=Dx^2-y^2-z^2\qquad  \longrightarrow \qquad Q_2=x^2-y^2-z^2.$$ It  induces  a change in the image of \eqref{isom1} and now the isomorphism  between
$\text{\rm PSL}_2(\R)=\text{\rm SL}_2(\R)/\{\pm I\}$ and the proper Lorentz group preserving $Q_{2D}=Dx^2-y^2-z^2$, except for a global change of sign, becomes
\[
 \begin{pmatrix}
  a&b \\ c&d
 \end{pmatrix}
 \longmapsto
 \begin{pmatrix}
  \frac{1}{2} \big( a^{2} +  b^{2} + c^{2} + d^{2}\big)
  &
  \frac{1}{2\sqrt{D}}\big(-a^{2} + b^{2} - c^{2} + d^{2}\big)
  &
  -\frac{1}{\sqrt{D}}(a b + c d)
  \\
   \frac{\sqrt{D}}{2} \big(-a^{2} - b^{2} + c^{2} + d^{2}\big)
   &
   \frac{1}{2} \big(a^{2} - b^{2} - c^{2} + d^{2}\big)
   &
   a b - c d
   \\
   -(a c + b d) \sqrt{D}
   & a c - b d
   & b c + a d
 \end{pmatrix}.
\]
The crucial question is again how to characterize $a,\ b,\, c,\, d$ in the initial matrix having an integral matrix as image.
A simplification is reached, re-parametrizing the initial matrix as
\begin{equation}\label{mmatrix}
 M=
 \begin{pmatrix}
  -a-c\sqrt{D}    &   -b-d\sqrt{D}
  \\
  b-d\sqrt{D}    &   -a+c\sqrt{D}
 \end{pmatrix},
\end{equation}
because it rules out the square roots in the image, namely, the previous isomorphism becomes
\begin{equation}\label{isom2}
 F\,:\,
 M
 \longmapsto
 \begin{pmatrix}
  a^{2} + b^{2}+D(c^2+d^2) & -2 a c + 2 b d & -2 b c - 2 a d
  \\
  -2 D( a c + b d) & a^{2} - b^{2} +D(c^2-d^2) & 2ab+2  D c d
  \\
  2 D (b c - a d) & - 2 a b+2 D c d  & a^{2} - b^{2}-D(c^2-d^2)
 \end{pmatrix}.
\end{equation}
In these terms it is plain to see that for $a,b,c,d\in\Z$ it is obtained an integral matrix. In fact, as all the entries are quadratic, one can expect that it is possible to introduce $\sqrt{2}$ denominators in $M$ if the parity conditions cancel the $1/2$ in the diagonal entries of the image, as above. Elaborating this idea, it can be proved \cite{FrKl} (cf. \cite{magnus}) that the case $4\mid D-1$ parallels the case $D=1$. Namely, the group is composed by two parts that reduce to $\Gamma_\theta$ and the companion coset $C_\theta$ for $D=1$. Specifically, 
\[
 \Gamma=\Gamma'_\theta\cup C'_\theta
\]
where
\[
 \Gamma'_\theta
 =
 \big\{M\in \text{\rm PSL}_2(\R) \,:\, a,b,c,d\in \Z\big\}
 \qquad\text{and}\qquad
 C'_\theta
 =
 \Big\{\frac{1}{\sqrt{2}}M\in \text{\rm PSL}_2(\R) \,:\, a,b,c,d\in \Z\Big\}.
\]
Here $M$ is like in \eqref{mmatrix}.

If instead $4\nmid D-1$ then it is necessary to add a new set, which is a kind of semi-integral version of $\Gamma'_\theta$ namely
$$
 \Gamma=\Gamma'_\theta\cup C'_\theta\cup C''_\theta,
$$
where $\Gamma'_\theta$ and $C'_\theta$ are as before and
$$
 C''_\theta
 =
 \Big\{\frac{1}2 M\in \text{\rm PSL}_2(\R) \,:\, a,b,c,d\text{ odd integers}\Big\}.
$$
In fact, the distinction between $4\mid D-1$ and $4\nmid D-1$ is a little artificial (only to emphasize the analogy of the former case with $D=1$) because $C''_\theta$ is empty for $4\mid D-1$. It is also empty for $D$ even because, by computing the determinant, 
$\frac 12 M\in\text{\rm PSL}_2(\R)$ implies $\frac 12(a^2+b^2)=2+\frac 12 D(c^2+d^2)$. If $a,\,b,\,c,\,d$ are odd then the left side is odd and the right side is even.
\medskip 

In any case, \eqref{sol2d} with $F$ given by \eqref{isom2} produces infinitely many solutions of $Dx^2-y^2-z^2=q^2$ showing that $\mathcal{N}_0$ is infinite. As in the one dimensional case, the theory assures \cite[Lemma\,6.1]{cassels} that there is a finite number of infinite families of solutions. That is to say, with a finite number of choices of $(s,k_0,\ell_0)$ all the needed solutions to construct $\mathcal{N}_0$ are reached with \eqref{sol2d}.
\medskip

Let us close this section with some comments about $q^2\in\Q-\Z$. Instead of deepening into the theory, we only show a shortcut to deduce $\#\mathcal{N}_0=\infty$ from our previous knowledge about $q^2\in\Z$.

If $q^2\in\Q-\Z$, \eqref{sqf2} must be replaced by 
$$
  k_0^2+\ell_0^2+q^2=\frac{Ds^2}{D_*s_*^2}
 \qquad\text{with $D$ and $D_*$ squarefree and coprime}.
$$
Clearing the denominator and recalling \eqref{cond}, it is deduced $$D\bigg(\frac{a_js}{b_j}\bigg)^2-D_*s_*^2k_j^2-D_*s_*^2\ell_j^2=q^2s_*^2.$$ Hence
$$
  \mathcal{N}_0
  =
  \big\{
  (y,z)\,:\, Dx^2-D_*s_*^2y^2-D_*s_*^2z^2=q^2s_*^2\text{ with }(x,y,z)\in\Z^3
  \big\}.
$$
In principle, one could study the group leaving invariant the quadratic form 
$$Q_{2 D D_\ast }=Dx^2-D_*s_*^2y^2-D_*s_*^2z^2,$$ and proceed as before \cite{FrKl}, but some technical issues appear because $D_*s_*^2$ is not squarefree if $s_*>1$. A trick to overcome this problem is to re-write the equation as 
\[
 DD_* x^2-(D_*s_*y)^2-(D_*s_*z)^2=q^2s_*^2.
\]
Then one has to look for integral solutions of $DD_*x^2-Y^2-Z^2=q^2s_*^2$ and furthermore to restrict  $Y$ and $Z$ to be multiples of $D_*s_*$. Note that $DD_*$ is squarefree and therefore the action
\eqref{sol2d} with $F$ obtained from \eqref{isom2} with the replacement $D \to DD_*$ and $(s,k_0,\ell_0)\to (s,D_*s_*k_0,D_*s_*\ell_0)$ produces infinitely many solutions $(x,Y,Z)$. The main issue is to prove that infinitely many of them verify that their two last coordinates are multiples of $D_*s_*$. This can be done as follows.  The $3\times 3$ matrices of determinant~$1$ modulo $D_*s_*$ form a finite group, and this implies that there exists $N\in\Z^+$ such that
\[
 \big(F(\gamma)\big)^N=F(\gamma^N)
 \equiv \text{Id}\mod{D_*s_*}.
\]
Hence
\[
  F(\gamma^N)
 \begin{pmatrix}
  s\\ D_*s_*k_0\\ D_*s_*\ell_0
 \end{pmatrix}
 \equiv 
 \begin{pmatrix}
  s\\ 0\\ 0
 \end{pmatrix}
 \mod{D_*s_*}.
\]
As $\{\gamma^N\,:\, \gamma\in\Gamma\}$ is infinite, there are infinite solutions with $D_*s_*\mid Y,Z$ and $\mathcal{N}_0$ is infinite too. 

Although it is possible to obtain an explicit formula for $N$, for moderate values of $D_*s_*$  may be more useful in practice to compute the image of successive powers of $\gamma$  until its image by $F$ fulfills the divisibility conditions.

\section{Some numerical simulations}

The purpose of this section is to explain some numerical examples in detail and to plot some quantum carpets\footnote{The Matlab/Octave code generating the figures and the full size color images are posted on the web page \url{https://matematicas.uam.es/~fernando.chamizo/dark/d_talbot.html} along with more related material.}. The latter in the 1D case corresponds to density plots of 
$|\Psi(\phi_2,t)|$, the square root of the probability density for the state in \eqref{state} with $t$ in the horizontal axis and $\phi_2$ in the vertical axis. The only reason to prefer $|\Psi|$ instead of the more natural $|\Psi|^2$ is to avoid the saturation giving less appealing images.
These plots involve a choice of the coefficients $\lambda_{\vec{n}}$ indicated in each case. The angle is considered in the range $[-\pi,\pi]$, as mentioned in the first section. On the other hand, the natural timel scale is $T_{\text{rev}}$, the revival time. Hence $[0,1]$ in the plot must be interpreted as $[0,T_{\text{rev}}]$ in usual units.  In the 2D case a density plot of $|\Psi(\phi_1,\phi_2,t)|$ is not possible because it would require three dimensions. It will replaced below by plots for close fixed values of $t$ to exemplify the evolution of the state.

\

The first example corresponds to a 1D case  having a finite $\mathcal{N}_0$, corresponding to the values $\ell_0=3$ and $q=\frac 32 \sqrt{21}$. Then $\ell_0^2+q^2= \frac{15^2}{2^2}$ which corresponds in \eqref{n01dr} to $D=D_*=1$, $s_*=3$. Using the explicit formula for $\mathcal{N}_0$ in terms of the divisors of $q^2s_*^2=189$, it is obtained 
\[
 \mathcal{N}_0
 =
 \big\{
 (0,\pm3), 
 (0,\pm5), 
 (0,\pm15), 
 (0,\pm47) 
 \big\}.
\]
Here all the divisors of $189$ satisfy the extra condition of $\frac{189}{d}-d$ being multiple of $4$. By computing $$\frac{a_j}{b_j}=\frac{\sqrt{\ell_j^2+q^2}}{\sqrt{\ell_0^2+q^2}},$$ 
where the last identity follows according to \eqref{1dcond}, it is obtained that
 $$\frac{a_1}{b_1}=1,\qquad  \frac{a_2}{b_2}=\frac{17}{15},\qquad  \frac{a_3}{b_3}=\frac{11}{5},\qquad \frac{a_4}{b_4}= \frac{19}{3}.$$  Hence $L=15$ and the revival time is $T_{\text{rev}}=\frac{30\pi}{\omega_{\vec{n}_0}}$. This time remains invariant when $(0,\pm 47)$ is omitted. The  first two figures are the plots of the corresponding quantum carpets in these situations with $1$ coefficients.

\begin{center}
 \begin{tabular}{c}
  \includegraphics[height=160pt]{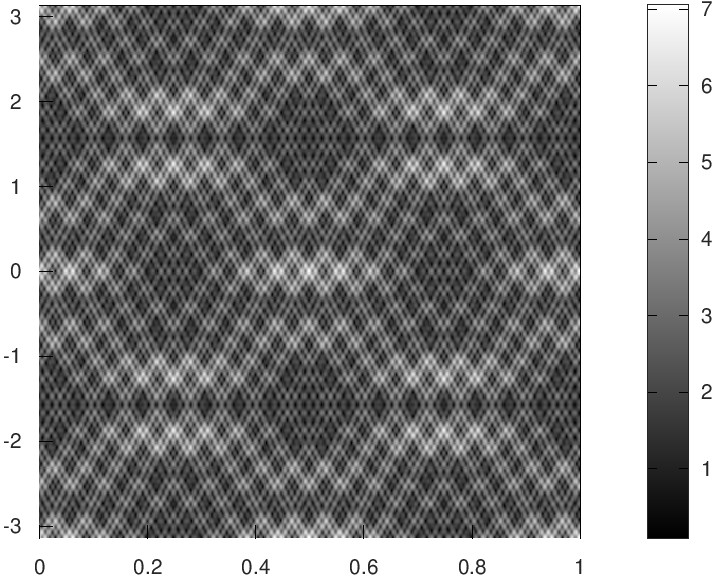}
  \\
  {\sc Fig.\,1.} {\sf Full $\mathcal{N}_0$ with $\lambda_{\vec{n}}=1$.}
 \end{tabular}
 \quad
 \begin{tabular}{c}
  \includegraphics[height=160pt]{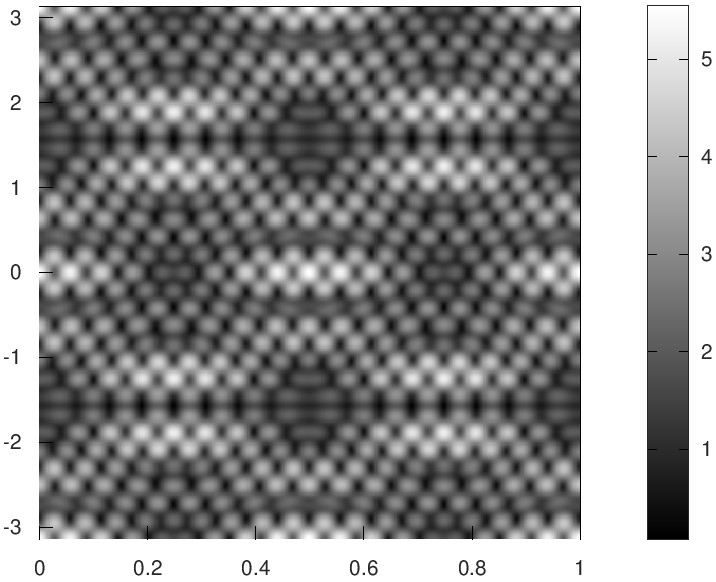}
  \\
  {\sc Fig.\,2.} {\sf $\mathcal{N}_0-\{(0,\pm 47)\}$ with $\lambda_{\vec{n}}=1$.}
 \end{tabular}
\end{center}

\

As a second example, the values $\ell_0=5$ and $q=\sqrt{2}$ will be considered, in order to show how to obtain the result claimed in \eqref{example1d}. Recall that in this situation $s=D=3$. It was already shown in previous sections, that there is only a family of solutions, so 
$\mathcal{N}_0=\{(0,x_n)\,:\, n\in\Z\}$ where, according to \eqref{gpf}
\[
 x_n+y_n\sqrt{3}
 =
 \pm (5+3\sqrt{3})(2+\sqrt{3})^n.
\]
Taking apart the $\pm$, this recurrence produces 
$x_{n+1}+y_{n+1}\sqrt{3}=(2+\sqrt{3})(x_n+y_n\sqrt{3})$, a procedure leading to the first order recurrence
\begin{equation}\label{nina}
 (x_{n+1},y_{n+1})
 =
 (2x_n+3y_n,x_n+2y_n).
\end{equation}
By making a further recurrence step and by eliminating $y_n$ and $y_{n+1}$, it is obtained the second order recurrence
\[
 x_{n+2}=4x_{n+1}-x_{n}
 \qquad\text{with}\quad 
 x_0=5,\ x_1=19\quad\text{for}\quad n\in\Z.
\]
It is clear that $x_{-3-n}=-x_{-n}$ for $n=0,1$, and this identity generalizes for $n$. Hence, by defining $c_n=x_{n-1}$ for $n\ge 0$, the formula \eqref{example1d} follows. In the same way, $c_n'=y_{n-1}$ satisfies $c'_{n+2}=4c'_{n+1}-c'_{n}$ with starting values $c_0'=1$, $c_1'=2$.
An easy inductive argument using the recurrence shows that $c_n'$ and $c_{n+1}'$ are coprime (in particular, at least one is not divisible by~$3$). Therefore for any $\mathcal{N}$ containing $(0,c_n)$ and $(0, c_{n+1})$ for some $n$ we have $L=3$ and  the revival time is 
\[
  T_{\rm rev}
  =
  \frac{2\pi R}{c\sqrt{3}}
  =
  \frac{6\pi }{\omega_{\vec{n}_0}}.
\]

Due to the exponential growth of $c_n$, even for fairly small cardinalities of $\mathcal{N}$, the state \eqref{state} shows large variations for tiny changes of $\phi_2$ and $t$ and the density plot is close to be a cloud of random points. In this context it is natural to let $\lambda_{\vec{n}}$ to decay with $|\vec{n}|$. In Fig.~3 and Fig.~4 there are examples choosing $\ell_j$ the smallest values of $c_n$ and 
$\lambda_{\vec{n}}=|\vec{n}|^{-1/4}=|\ell_j|^{-1/4}$.
\begin{center}
 \begin{tabular}{c}
  \includegraphics[height=160pt]{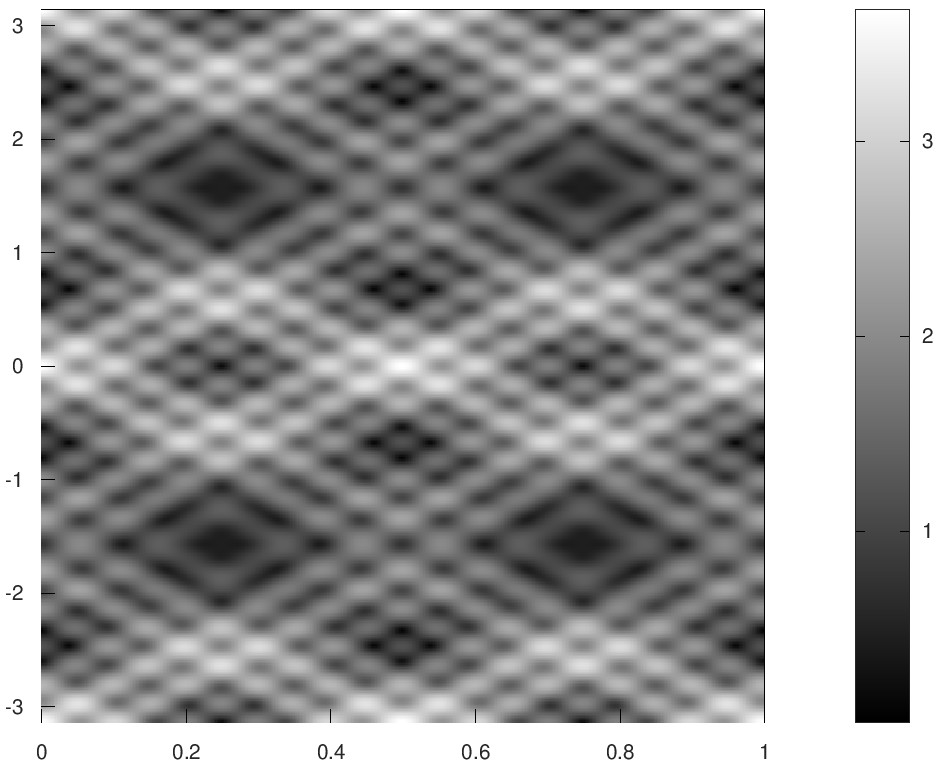}
  \\
  {\sc Fig.\,3.} {\sf $\ell_j=\pm1,\pm5,\pm19$, $\lambda_{\vec{n}}=|\ell_j|^{-1/4}$.}
 \end{tabular}
 \quad
 \begin{tabular}{c}
  \includegraphics[height=160pt]{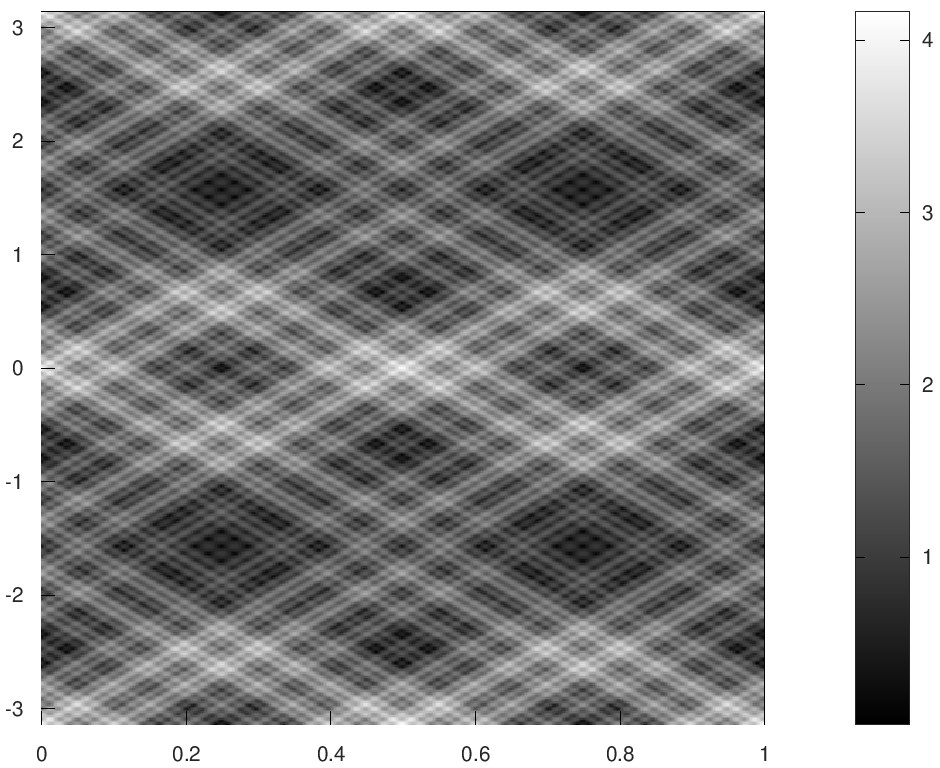}
  \\
  {\sc Fig.\,4.} {\sf $\ell_j=\pm1,\pm5,\pm19,\pm71$, $\lambda_{\vec{n}}=|\ell_j|^{-1/4}$.}
 \end{tabular}
\end{center}
As a matter of fact, for each $t$ fixed, when $\#\mathcal{N}\to\infty$ it is obtained a lacunary series and according to known facts in fractal geometry \cite[\S11.3]{falconer2} the graphs of the real and imaginary part give rise to a fractal, a curve with fractional box dimension. 

In this and other examples, it is apparent an underlying structure based on interwoven straight lines with 2 different slopes. The explanation is that in \eqref{state}, due to the exponential growth, $\frac{\sqrt{\vec{n}^2+q^2}}{n_2}$ is very early close to the sign of $n_2$ and the pairs of lines correspond to the light cones. In fact if all the $\ell_j$ are chosen to be positive, commonly the density plot looks like a dull collection of oblique parallel bands. 

\

In order to mask the light cones, a possibility is to take $q^2$ large in order to avoid $$\frac{\sqrt{\ell_j^2+q^2}}{|\ell_j|} \sim~1,$$ at least for the first values of $\ell_j$. With this idea in mind and to illustrate the appearance of several families of solutions, consider the example
\[
 \vec{n}_0=(0,3)
 \qquad\text{with}\qquad 
 q=\sqrt{791}. 
\]
From \eqref{sqf}, $D=2$ and $s=20$. Consequently, the generalized Pell's equation is $x^2-2y^2=-791$ and by \eqref{gpf}, it is seen that it admits a family of solutions described by 
\[
  x+y\sqrt{2}
  =
  \pm
  \big(3+20\sqrt{2}\big)
  \big(3+2\sqrt{2}\big)^n
  \qquad\text{with}\quad n\in\Z
\]
because $(3,2)$ is a minimal positive solution of $x^2-2y^2=1$. However, solutions $(x,y)$ in other families may exist. The condition $0<2a<791(1+2\sqrt{2\cdot 3})$ for a positive solution $(a,b)$ to be the seed of a new family, by direct calculations, only gives $(19,24)$. So ii is obtained  the second family 
\[
  x+y\sqrt{2}
  =
  \pm
  \big(19+24\sqrt{2}\big)
  \big(3+2\sqrt{2}\big)^n
  \qquad\text{with}\quad n\in\Z
\]
and any solution of $x^2-2y^2=-791$ belongs to one of them. 

Taking $n\ge 0$, the first positive values for $x$ in the first family are 
\[
 3,89, 531, 3097, 18051, 105209, 613203,\dots
\]
which satisfy the recurrence $c_{k+1}=6c_k-c_{k-1}$. For $n<0$ the values are 
\[
 71, 429, 2503, 14589, 85031, 495597, 2888551,\dots
\]
and, naturally,  they satisfy the same recurrence with different starting values. In the same way, for the second family it is obtained for $n\ge0$
\[
 19, 153, 899, 5241, 30547, 178041, 1037699,\dots
\]
and for $n<0$
\[
 39, 253, 1479, 8621, 50247, 292861, 1706919,\dots
\]
still satisfying the same recurrence. Then 
\[
 \mathcal{N}_0
 =
 \big\{
 (0,\pm c_k^{(j)})\,:\, 1\le j\le 4,\  k\in\Z_{\ge 0}
 \big\}
\]
where $c_{k+1}^{(j)}=6c_k^{(j)}-c_{k-1}^{(j)}$ with the starting values 
\[
 \begin{array}{r|c|c|c|c|}
  &j=1&j=2&j=3&j=4
  \\
  \hline
  (c_k^{(0)}, c_k^{(1)})
  & (3,89)
  & (71,429)
  & (19,153)
  & (39,253)
  \\
  \hline
 \end{array}
\]

In Fig.~5 it is considered the subset 
\[
 \mathcal{N}=\big\{(0,\pm 3),(0,\pm 19),(0,\pm 39)\big\}
\]
with the coefficients $\lambda_{\vec{n}}=|\vec{n}|^{-1/4}$, as before, and in Fig.~6 it is added $(0,\pm 71)$.
In both cases the revival time is $T_{\text{\rm rev}}=\frac{20\pi}{\omega_{\vec{n}_0}}$ because the $\frac{a_j}{b_j}$ are $1$, $\frac{6}{5}$, $\frac{17}{10}$ and $(0,\pm 71)$ adds $\frac{27}{10}$ that does not change the value of $L$ which is~$10$. 
\begin{center}
 \begin{tabular}{c}
  \includegraphics[height=160pt]{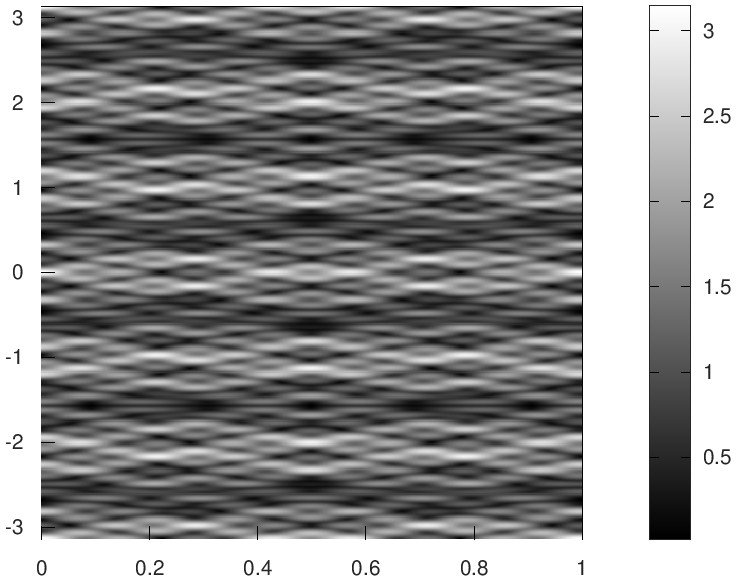}
  \\
  {\sc Fig.\,5.} {\sf $\ell_j=\pm3,\pm19,\pm39$, $q^2=791$.}
 \end{tabular}
 \begin{tabular}{c}
  \includegraphics[height=160pt]{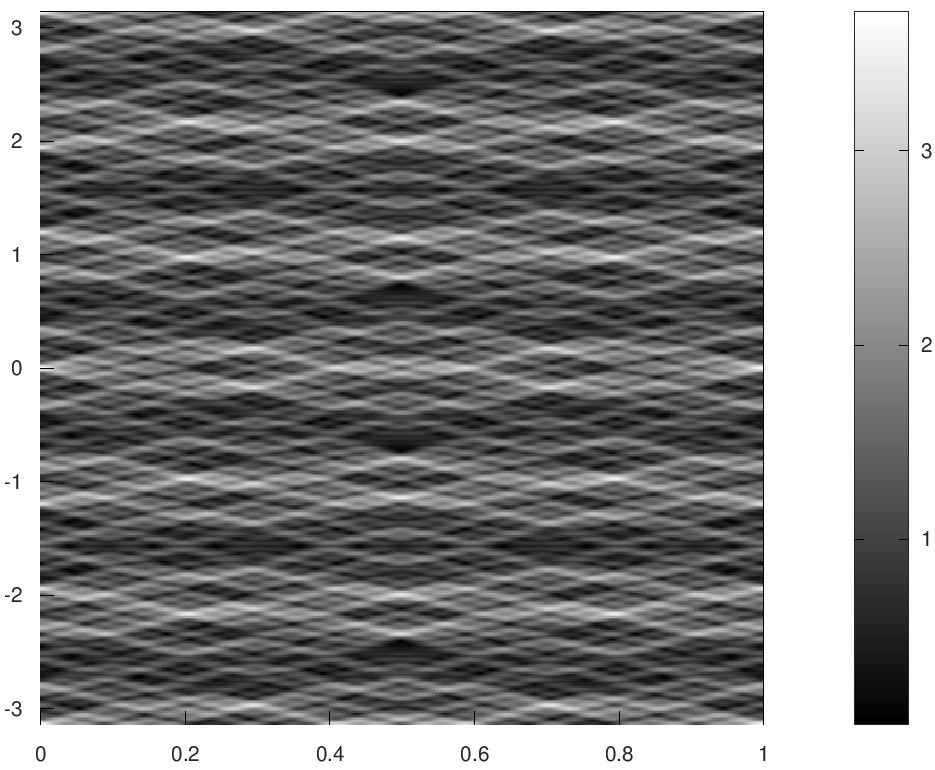}
  \\
  {\sc Fig.\,6.} {\sf $\ell_j=\pm3,\pm19,\pm39,\pm71$, $q^2=791$.}
 \end{tabular}
\end{center}
Note that, as expected, the clear line structure in the previous example coming from the light cones has been blurred, especially in the first figure. 

\

The examples given above correspond to a 1D case. It is a good point to exemplify the two dimensional situation.
Recall that in the 2D setting a way to escape from the condition $q^2\in\Q$ keeping $\mathcal{N}_0$ nontrivial was to consider numbers with many representations as a sum of two squares. 
According to the standard theory \cite[\S16.9]{HaWr}, if $n$ is a product of $k$ distinct primes of the form $4m+1$ then the number of representations of $n$ as a sum of two squares is $2^{k+2}$. By choosing $k_0^2+\ell_0^2=n$ it follows that $\#\mathcal{N}_0=2^{k+2}$ for any $q^2\not\in\Q$ while its energy set reduces to the single element $E=\frac{c\hbar \sqrt{n+q^2}}{R}$. Many representations are obtained applying the following~8 obvious symmetries
\[
 (n_1,n_2)\mapsto (\pm n_1,\pm n_2)
 \qquad\text{and}\qquad 
 (n_1,n_2)\mapsto (\pm n_2,\pm n_1).
\]
For instance, for $n=1105=5\cdot 13\cdot 17$, the $2^{3+2}=32$ representations come from applying the symmetries to the~4 pairs
\[
 (4,33),\quad
 (9,32),\quad
 (12,31),\quad
 (23,24).
\]
The value of $q^2$ does not affect to $\mathcal{N}_0$ and letting it grow, only the first coordinate in \eqref{estate} is relevant, giving in the limit 
\[
 \sum\Psi^2_{(\pm n_1,\pm n_2)}
 +
 \sum\Psi^2_{(\pm n_2,\pm n_1)}
 =
 4e^{-\frac{i Et}{\hbar}}\big(
 \cos(n_1\phi_1)\cos(n_2\phi_2)
 +
 \cos(n_2\phi_1)\cos(n_1\phi_2)
 \big),
\]
where the sum is over all sign combinations. 
Although this may seem very simple, the chaotic distribution of $n_1$ and $n_2$ for $n$ large leads to some hard questions in mathematical physics. Namely, even in this toral situation, some general conjectures \cite{berry_nodal} about the geometry of the \emph{nodal lines} (the zero level sets) of the states corresponding to the a given energy remain unsolved. 

To illustrate the richness of the situation, in Fig.~7 there is a plot of the nodal lines of the first coordinate of the state \eqref{state} for $\mathcal{N}_0$ as in the previous example with $k_0^2+\ell_0^2=1105$ and $\lambda_{\vec{n}}=1$ in the range
 $\phi_1,\phi_2\in [-\frac{\pi}{2},\frac{\pi}{2}]$. 
\begin{center}
 \begin{tabular}{c}
  \includegraphics[height=160pt]{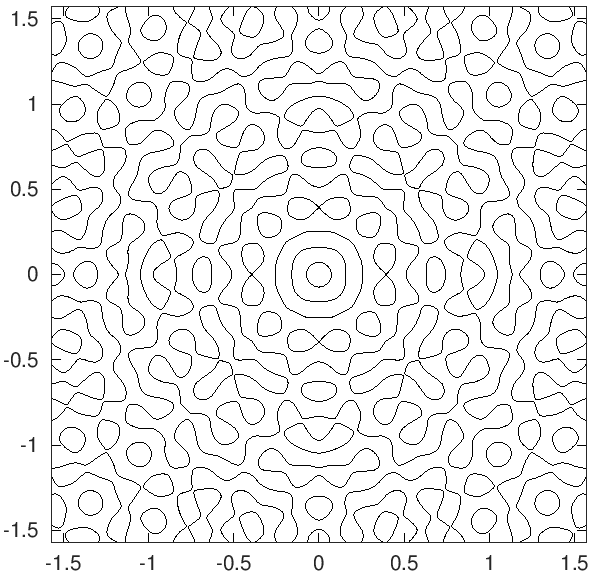}
  \\
  {\sc Fig.\,7.} {\sf Nodal lines $k_0^2+\ell_0^2=1105$, $\lambda_{\vec{n}}=1$.}
 \end{tabular}
 \begin{tabular}{c}
  \includegraphics[height=160pt]{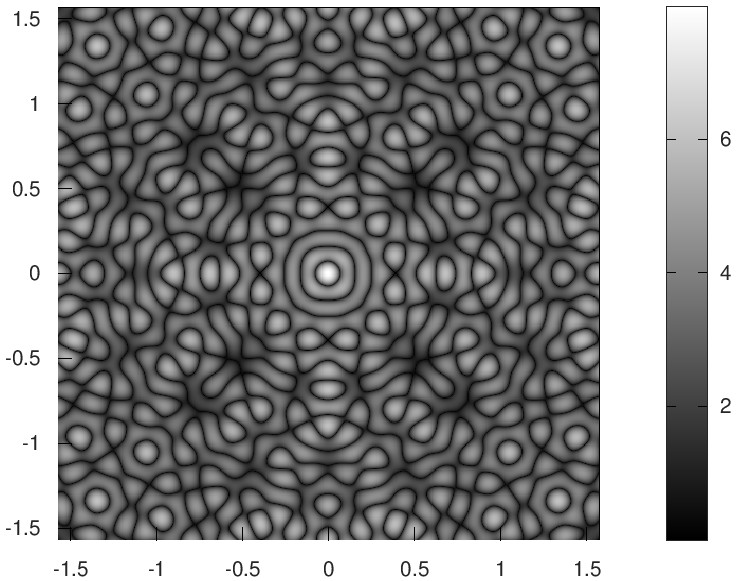}
  \\
  {\sc Fig.\,8.} {\sf $|\Psi(\phi_1,\phi_2)|^{0.5}$, $\lambda_{\vec{n}}=1$.}
 \end{tabular}
\end{center}

The image in Fig.~8 is the corresponding density plot of $|\Psi|^{0.5}$. As $e^{-iEt/\hbar}$ is constant, $|\Psi|$ does not depend on $t$.  The exponent $0.5$ is only for visualization purposes. It has been introduced to mollify the effect of the peak at $\phi_1=\phi_2=0$.

\

Consider now the example $(k_0,\ell_0)=(1,2)$ with $q^2=3$. The decomposition \eqref{sqf2} gives $D=s=2$ and $\mathcal{N}_0$ is composed by the $(y,z)$ pairs got from the solutions of $2x^2-y^2-z^2=3$. If one only wants to compute a small subset of $\mathcal{N}_0$ then one can proceed by direct inspection instead of using the parametrization by the Fuchsian groups. Let us take
\[
 \mathcal{N}=
 \big\{
 (\pm1,\pm2),\,
 (\pm2,\pm1),\,
 (\pm2,\pm5),\,
 (\pm5,\pm2),\,
 (\pm2,\pm11),\,
 (\pm11,\pm2),\,
 (\pm5,\pm10),\,
 (\pm10,\pm5)
 \big\}
\]
which comes from $\big\{(1,2),\,(2,5),\,(2,11),\,(5,10)\big\}$ applying the symmetries. 
The resulting sequence
$\frac{a_j}{b_j}=\sqrt{\frac{(k_j^2+\ell_j^2+3)}{8}}$
is 1, 2, 4, 4. 
All the denominators are~1, then the revival time is 
$T_{\text{\rm rev}}=\frac{2\pi}{\omega_{\vec{n}_0}}$.

As mentioned before, the density plot of the associated state $\Psi$ is not feasible because there are three variables $(\phi_1,\phi_2,t)$. To illustrate the situation, Fig.~9 is composed by~$4$ images showing the density plots in $(\phi_1,\phi_2)$ for $t=0,\, \frac{T_{\text{\rm rev}}}{10},\, \frac{T_{\text{\rm rev}}}{5}$ and $\frac{T_{\text{\rm rev}}}{10}$.
\begin{center}
 \begin{tabular}{c}
  \includegraphics[height=90pt]{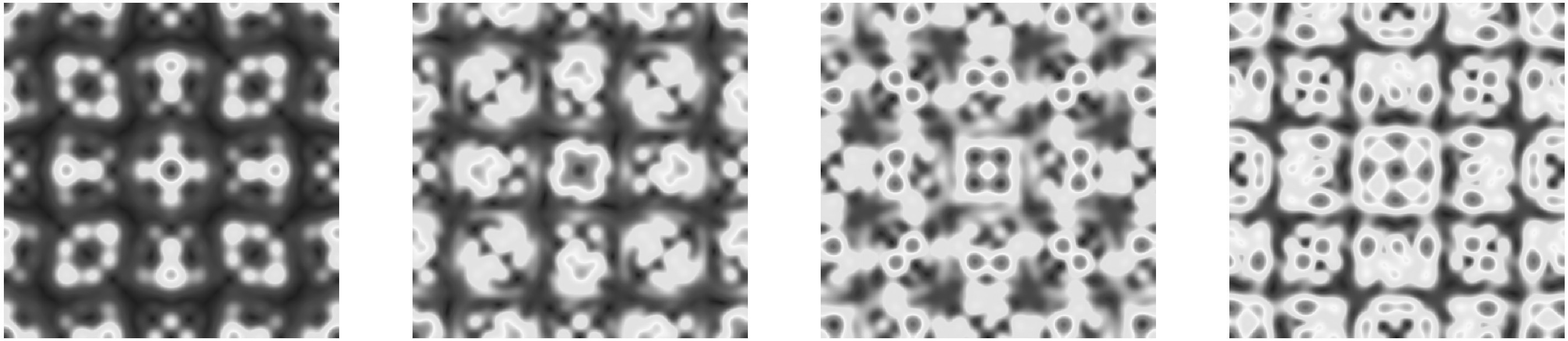}
  \\
  {\sc Fig.\,9.} {\sf $|\Phi(\phi_1,\phi_2, \frac{kT_{\text{\rm rev}}}{10})|$
  for $k=0,1,2,3$, $\phi_1,\phi_2\in [-\pi,\pi]$, with $\lambda_{\vec{n}}=1$.}
 \end{tabular}
\end{center}
Playing with different values, it seems that the sensitivity in $t$ is not uniform. For instance, the density plots for $t=0$ and $t=\frac{T_{\text{\rm rev}}}{20}$ are quite similar while there are noticeable differences between $t=\frac{T_{\text{\rm rev}}}{2}$ and $t=\frac{T_{\text{\rm rev}}}{2}-\frac{T_{\text{\rm rev}}}{20}$. Probably part of the explanation is the large peak corresponding to $\phi_1=\phi_2=t=0$ which masks relative differences.

\section{Discussion of the results}

In the present paper, the possible revivals for a relativistic free fermion ruled by the Dirac equation in a one dimensional and a two dimensional torus, were characterized.  
This was achieved by use of arithmetic tools such as Pell's equations or Fuchsian groups.
The revivals shown here are exact, that is, there is no need to take relativistic limit in order to detect them. These results generalize those found in \cite{strange} pointing revivals related to Pythagorean triples. In the present
paper, all the possible states exhibiting revivals were characterized. The interesting feature of these revivals is that they are known to hold for massless wave equations or for the standard Schr\"odinger equation. However
in the present context the dispersion relation is modified to $E=\sqrt{m^2+p^2}$ and still the revivals are exact. Although the torus topology may seem non realistic at first sight, it should be remarked that
the Dirac equation in several topologies may have applications in solid state physics. The study of revivals in this context is encouraging and perhaps may lead to interesting experimental results such as the ones reported in  \cite{ToRo}.
Finally, it may be possible and interesting to generalize the presented results to 3D, when the spatial coordinates are periodic. However it is likely that the analysis will be more involved. This could be an interesting task to be studied further.

\section*{Acknowledgments}
O.P. S. is grateful to the Universidad Autónoma and to the ICMAT de Madrid, where this work was performed, by their hospitality. The present work is supported by CONICET, Argentina and by the Grant PICT 2020-02181. This project has received funding from the European Union’s Horizon 2020 research and innovation program under the Marie Sk{\l}odowska-Curie grant agreement No 777822.
F. Ch. is partially supported by the PID2020-113350GB-I00 grant of the MICINN (Spain) and
by ``Severo Ochoa Programme for Centres of Excellence in R{\&}D'' (CEX2019-000904-S).


\end{document}